\documentclass[conference]{IEEEtran}

\usepackage{fancyhdr}
\usepackage[bookmarks=true,breaklinks=true,letterpaper=true,colorlinks,citecolor=blue,linkcolor=blue,urlcolor=blue]{hyperref}
\usepackage{cite}
\usepackage{url}
\usepackage{balance}
\usepackage{booktabs}
\usepackage{amsmath,amssymb,amsfonts}
\usepackage{algorithmic}
\usepackage{algorithm}
\usepackage{verbatim}
\usepackage{multirow}
\usepackage{bbding}
\usepackage{bm}
\usepackage{graphicx}
\usepackage{textcomp}
\usepackage{xcolor}
\usepackage{pifont}
\usepackage{multirow}
\usepackage{textcomp,mathcomp}
\usepackage{mathcomp}
\title{Tasa: Thermal-aware 3D-Stacked Architecture Design with Bandwidth Sharing for LLM Inference
\vspace{-10pt}
}

\author{
    \IEEEauthorblockN{
        Siyuan He\textsuperscript{1}, 
        Peiran Yan\textsuperscript{1,2}, 
        Yandong He\textsuperscript{1}, 
        Youwei Zhuo\textsuperscript{1}, 
        Tianyu Jia\textsuperscript{1,*}}
    \IEEEauthorblockA{
    \textsuperscript{1}\textit{School of Integrated Circuits} \\
    \textsuperscript{2}\textit{School of Software \& Microelectronics} \\
    \textit{Peking University}\\
    Beijing, China\\
    \textsuperscript{*}Corresponding Email: tianyuj@pku.edu.cn}
    \vspace{-20pt}
}
    
\begin{document}
\maketitle


\begin{abstract}

The autoregressive decoding in LLMs is the major inference bottleneck due to the memory-intensive operations and limited hardware bandwidth.
3D-stacked architecture is a promising solution with significantly improved memory bandwidth, which vertically stacked multi DRAM dies on top of logic die.
However, our experiments also show the 3D-stacked architecture faces severer thermal issues compared to 2D architecture, in terms of thermal temperature, gradient and scalability.
To better exploit the potential of 3D-stacked architecture, we present Tasa, a heterogeneous architecture with cross-stack thermal optimizations to balance the temperature distribution and maximize the performance under the thermal constraints.
High-performance core is designed for compute-intensive operations, while high-efficiency core is used for memory-intensive operators, e.g. attention layers.
Furthermore, we propose a bandwidth sharing scheduling to improve the bandwidth utilization in such heterogeneous architecture.
Extensive thermal experiments show that our Tasa architecture demonstrates greater scalability compared with the homogeneous 3D-stacked architecture, i.e. up to 5.55 $\tccentigrade$, 9.37 $\tccentigrade$, and 7.91 $\tccentigrade$ peak temperature reduction for 48, 60, and 72 core configurations.
Our experimental for Llama-65B and GPT-3 66B inferences also demonstrate 2.85 $\times$ and 2.21 $\times$ speedup are obtained over the GPU baselines and state-of-the-art heterogeneous PIM-based LLM accelerator.

\end{abstract}

\section{Introduction}

Large language models (LLMs), such as GPTs \cite{gpt} and Llama \cite{llama}, have exhibited extraordinary ability for diverse natural language processing applications, including conversational agents, text generation, summarization, etc.
LLM inference consists of two operation stages, i.e. prefill and decoding. 
In the prefill stage, the LLM takes the prompt tokens as input to generate the first output token.
After that, the LLM begins autoregressive decoding, using the last generated tokens and previous context as input to predict the next token in each iteration.
Prefill and decoding of LLMs pose stringent high requirements for computation performance and memory bandwidth, respectively.

Prior studies \cite{attacc_2024, specpim_2024, neupims_2024, duplex} show that the sequential autoregressive decoding is the latency bottleneck of LLM inference due to its memory-intensive characteristics and the limited memory bandwidth.
For example, NVIDIA A100 GPUs \cite{a100} have high computing performance (312 TFLOPS) but limited off-chip DRAM bandwidth (1555 GB/s).
To address the memory bandwidth bottleneck, processing-in-memory (PIM) technique is a promising approach, which integrates processing units near DRAM banks and utilizes the high internal bank-level memory bandwidth.
In recent years, DRAM vendors presented PIM designs with enhanced general matrix-vector multiplication (GEMV) performance, such as GDDR-PIM from SK Hynix \cite{gddr_pim}, HBM-PIM and LPDDR-PIM from Samsung \cite{hbm_pim,lpddr_pim,samsung_PIM}.
Furthermore, prior works \cite{attacc_2024, neupims_2024, specpim_2024} have explored GPU-PIM heterogeneous systems to optimize LLM inference, in which memory-intensive attention layers are offloaded to PIM while the compute-intensive fully connected layers are executed on GPU.
However, these architectures are only efficient for scenarios with high operation intensity (Op/B) and does not perform well across a wide range of Op/B.

The limitations of existing solutions motivate architecture explorations with higher memory bandwidth, such as 3D-stacked logic-to-DRAM architecture \cite{tetris_3d, alibaba_3d, asscc_3d, iscca24_3d}.
In NVIDIA's vision for future accelerator design, the 3D-stacked architecture will be employed to keep scale compute \cite{nvidia_3d}.
Such 3D-stacked architecture vertically stacks DRAM dies on top of a logic die via hybrid bonding (HB), where the DRAM and log dies utilize independent process technology.
The state-of-the-art stacking technology has also developed high-density hybrid bonding vias and fine pitch mini-TSVs \cite{sedram, sedram2} for higher IO parallelism and memory bandwidth.

\begin{figure}[t]
  \centering
  \includegraphics[width=0.95\linewidth]{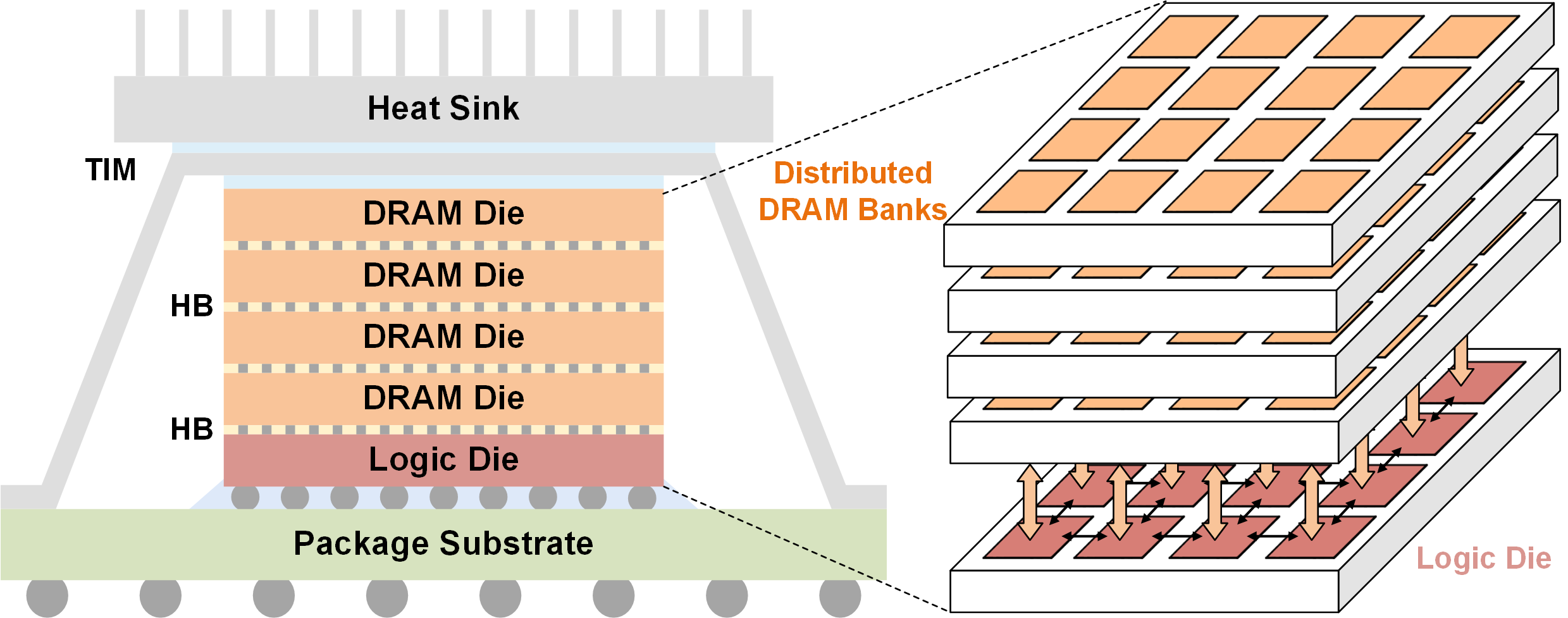}
  \vspace{-5pt}
  \caption{3D-stacked architecture with DRAMs on logic die.}
  \vspace{-15pt}
  \label{fig: 3d}
\end{figure}

Although 3D-stacked architecture has promising attributes, it is faced with serious thermal issues for applications such as LLM inferences.
When there are more dies in vertical direction, the power density will increase accordingly and the thermal issue will have a significant impact on reliability, efficiency, and longevity of 3D stacked chips.
As reported in \cite{thermal_failure}, thermal failures account for 55\% of all issues encountered in 3D stacked chips.
Based on our analysis, the thermal issue becomes more severe for high-performance designs with larger 3D stacked areas, significantly impacting the computing performance, i.e. similar to dark silicon \cite{dark_silicon}.
Due to these thermal issues, the 3D-stacked designs have restricted power constraints and require more careful architectural-level thermal considerations compared to 2D architectures.

In this work, we present Tasa, a 3D-stacked architecture design with thermal-aware engines for LLM inference.
To better understand the thermal challenges of LLM inference for 3D architecture, we first conduct a comprehensive simulation and analysis of the thermal distribution of 3D-stacked architecture with real-world LLM workloads.
Compared to conventional 2D architectures, we observe that 3D architectures exhibit both significantly higher overall temperature and greater variation in heat dissipation across different horizontal regions, making them more prone to localized thermal hot spots.
The distinct differences in power density among them create thermal optimization opportunities in the heterogeneous floorplan.
Therefore, we present our 3D-stacked architecture, Tasa, a thermal-aware 3D-stacked heterogeneous architecture with GEMM-oriented performance-cores (P-cores) and GEMV-specific efficiency-cores (E-cores).
By considering both the computational characteristics of LLM operators and its thermal impact, Tasa can balance the heat flow and achieve a more uniform temperature distribution, effectively lowering the overall temperature.
To further enhance the end-to-end performance, we implement a bandwidth-sharing scheduling algorithm and a hierarchical NoC architecture to leverage unused bandwidth from P-cores to E-cores, improving the bandwidth utilization and reducing the latency.
Experiment results show Tasa achieves superior thermal performance over homogeneous logic die by up to 5.55 $\tccentigrade$, 9.37 $\tccentigrade$, and 7.91 $\tccentigrade$ reduction under 48, 60, and 72 cores for Llama-65B and GPT-3 66B inference.
Compared with the GPU baseline and the state-of-the-art LLM accelerator, i.e. GPU-PIM, Tasa achieves average 2.85$\times$ and 2.21$\times$ performance improvements, while outperforms the homogeneous 3D-stacked architecture by 1.33$\times$.

The contributions of this work are summarized as follows:

\begin{itemize}
\item
We compare the thermal differences between 3D-stacked and 2D architectures and analyze the root causes of the distinct thermal distribution issue.

\item
We present a thermal-aware heterogeneous architecture with both P-cores and E-cores on logic die to alleviate thermal issues of different power density.

\item
We introduce a bandwidth-sharing scheduling algorithm and a hierarchical NoC for Tasa to enhance the bandwidth utilization and improve the performance.

\item
We evaluate Tasa and demonstrate that it presents significant thermal benefits and performance over GPU, PIM-GPU accelerator, and homogeneous 3D-stacked design.

\end{itemize}

\section{Background}

\subsection{LLMs and 3D Architectures}

Modern LLM inference consists of two distinct stages: the prefill and decoding.
During prefill, the model takes the prompt sequence as input and generates the first token and KV cache for decoding.
In the decoding stage, the model iteratively takes the last output token and previous KV cache as input to generate a new token and update the KV cache.
LLMs are structured based on Transformer blocks \cite{attention}, which consists of a multi-head attention (MHA) module and a feed-forward network (FFN), accompanied by layer normalization and residual operators. 
Due to the sequential nature of autoregressive decoding, the decoding accounts for most of the end-to-end inference latency.
For example, the decoding stage consumes more than 96\% of the total execution time in the GPT-3 175B model for 32 input tokens and output tokens \cite{attacc_2024}.

To enhance the LLM inference throughput, emerging 3D-stacked architecture with HB is a promising solution.
The architecture stacks multi-DRAM dies on top of a logic die with HB and mini-TSV technology.
The dummy HB is also employed across layers to reduce thermal stress and temperature gap \cite{asscc_3d}.
Fig. \ref{fig: 3d} shows a typical 3D design topped with thermal interface material (TIM) and a heat sink.
For 3D-stacked DRAM, the array blocks are distributed with fine granularity.
Each block is an independent memory channel and is connected to its local logic core by HB.
The high density HB via (110,00/$mm^2$ with 3um pitch) and 1.5um fine pitch mini-TSV unlocks extremely high I/O parallelism.
The direct bonding process significantly reduces parasitic resistance and capacitance, resulting in much lower logic-to-memory interface power consumption than HBM.
For example, SeDRAM \cite{sedram} is 3D connected to logic die through HB and achieves 34GB ps/Gbit bandwidth and 0.88 pJ/bit efficiency, which is 6.2$\times$ higher bandwidth and 4.35$\times$ better efficiency than HBM3 \cite{hbm3}.
Prior works have exploited 3D-stacked architectures in various scenarios.
Niu et al. \cite{alibaba_3d} presented a 3D-stacked processing-near-memory engine for the memory-bound recommendation system.
Yue et al. \cite{iscca24_3d} accelerate vision AI models with 3D-stacked DRAM.

\subsection{Thermal Considerations for 3D ICs}

Thermal is always a challenging issue for high-performance computing, particularly in 3D chips. 
The vertically stacked dies not only enhance power density but also introduce higher thermal resistance.
The larger distance between logic die and the heat sink further diminishes the efficiency of the cooling effect, resulting in poor heat dissipation. 
These thermal limitations restrict the achievable peak performance and memory bandwidth, ultimately degrading the advantages of the 3D-stacked architecture.
To improve thermal dissipation, researchers have investigated various structures and materials,  alongside innovative cooling methods to address the heat dissipation in 3D chips \cite{thermal_review_1, thermal_review_2}.

In addition to the passive cooling solutions, diverse research has explored dynamic thermal managements (DTMs) based on runtime workloads \cite{thermal_cpu,thermal_dvfs,thermal_dvfs2}, such as dynamic voltage-frequency scaling (DVFS) and thread migration in CPUs. 
These thermal management schemes can dynamically regulate chip operation conditions and actively mitigate overheating and local hot spots.
Furthermore, a few studies also optimize the memory access and data storage layout in DRAM based on the memory access pattern \cite{thermal_dram0,thermal_dram1,thermal_dram2} to reduce DTM overhead.

For LLM inference workloads on 3D-stacked architecture, the thermal challenges are further intensified due to the high parallelism operators, necessitating a cross-stack optimization.
However, the prior works mostly focus on standalone area of either material, cooling, or DTM, while few works address thermal issues from the architectural perspective.
In fact, incorporating the thermal considerations at the design phase is crucial to unleash the potential of 3D-stacked architectures.
To enable effective thermal-aware architecture design and optimization, a comprehensive analysis of thermal features for 3D is essential.

\section{Thermal Analysis of 3D Architecture}

\begin{figure}[t]
  \centering
  \includegraphics[width=0.95\linewidth]{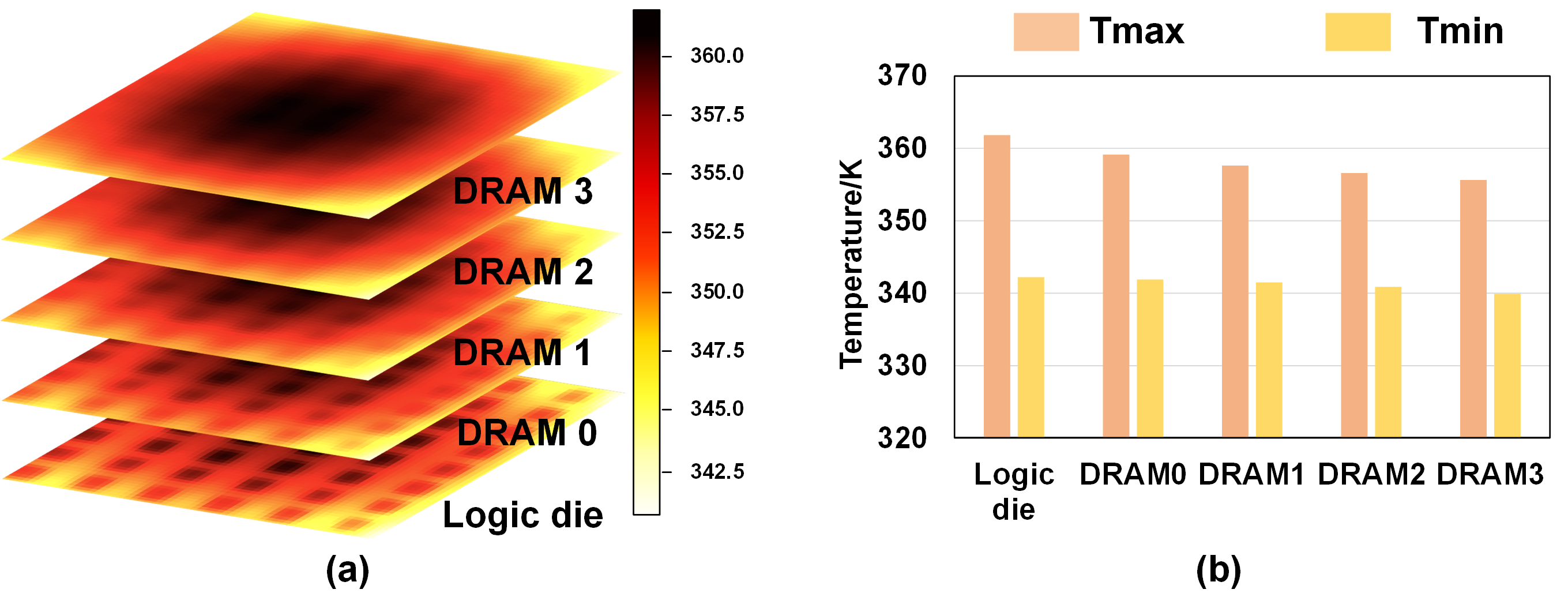}
  \vspace{-5pt}
  \caption{(a) The simulated heat map of 3D-stacked architecture for LLM inference, and (b) the temperature distribution across vertical dies.}
  \vspace{-10pt}
  \label{fig: heatmap}
\end{figure}

To well understand the thermal challenges for 3D-stacked architectures, we perform comprehensive profiling and analysis of temperature distribution in 3D architecture.
As shown in Fig. \ref{fig: heatmap}, we experiment with a 4-layer stacked memory with 8 TB/s bandwidth and a logic die which can be abstracted into memory, communication unit, and the computation array.
For the computation array, a GPU-like architecture \cite{a100} is adopted with 48 cores to achieve a performance of 196 TFLOPS, in which each core features two 32$\times$32 systolic arrays and occupies an area of 4mm$\times$2.5mm.
We evaluate the thermal profile for GPT-3 66B inference with a batch size of 32 using HotSpot 6.0 simulator \cite{HotSpot6_2015}. The detailed hardware simulation configurations are provided in Table \ref{tab: thermal}.

Fig. \ref{fig: heatmap} also shows steady-state temperature distributions in both horizontal and vertical directions.
It is observed that only a small temperature gradient (1-2 $\tccentigrade$) exists in the vertical direction across the stacked dies due to the dummy HB.
For example, the logic die has a 361.79 $\tccentigrade$ peak temperature while the stack DRAM dies have the temperature within a range of 355.66-359.11 $\tccentigrade$.
The horizontal temperature distribution is similar across layers, which is primarily influenced by the hot spots in the logic die  since the logic die has higher power density than the DRAM die.

\begin{figure}[t]
  \centering
  \includegraphics[width=0.95\linewidth]{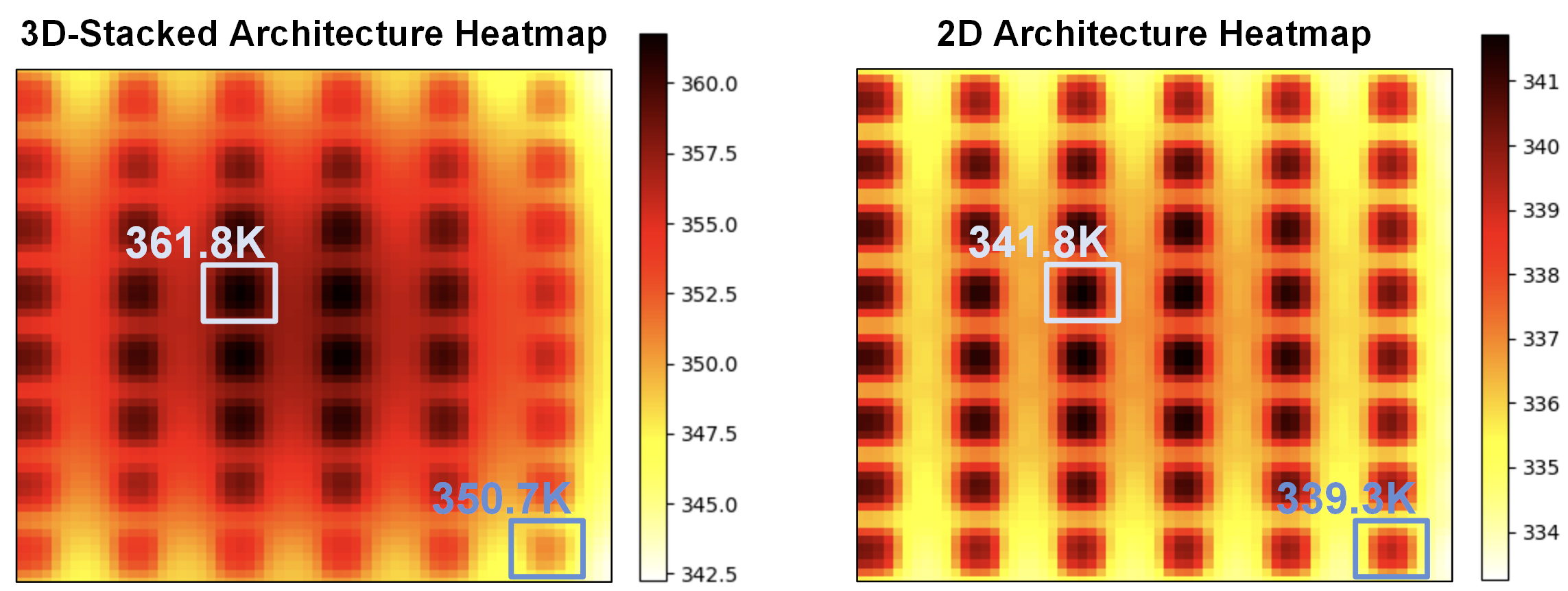}
  \vspace{-5pt}
  \caption{(a) The horizontal temperature distribution of the logic die in 3D architecture and (b) conventional 2D architecture.}
  \vspace{-15pt}
  \label{fig: hotspot_diff}
\end{figure}

\begin{figure}[t]
  \centering
  \includegraphics[width=0.95\linewidth]{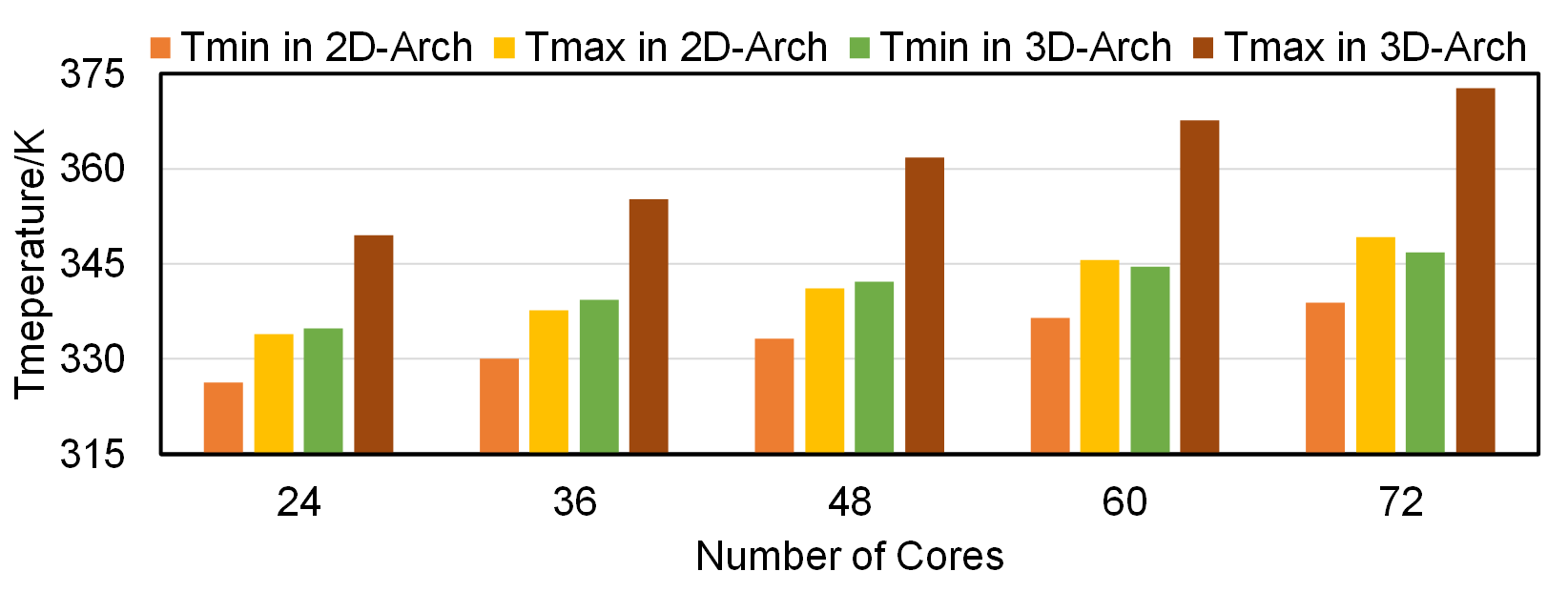}
  \vspace{-10pt}
  \caption{Simulated temperatures with varying cores for 2D/3D architectures. }
  \vspace{-10pt}
  \label{fig: t_diff}
\end{figure}

\begin{figure}[t]
  \centering
  \includegraphics[width=0.95\linewidth]{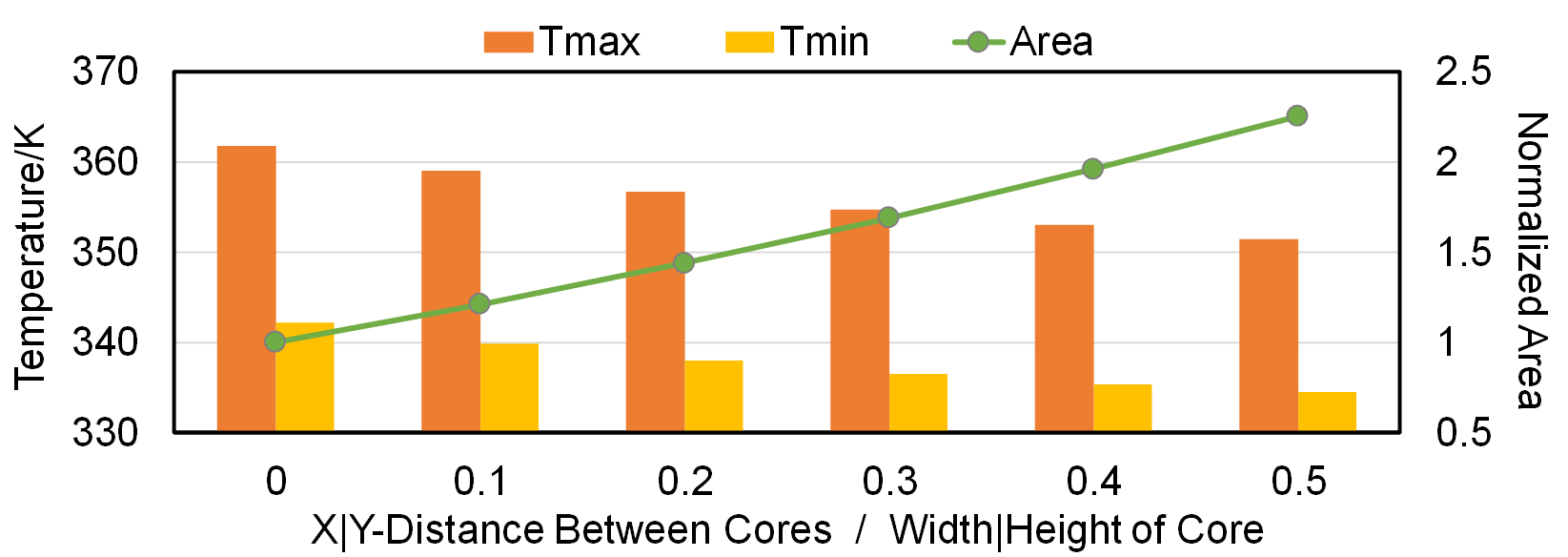}
  \vspace{-5pt}
  \caption{Thermal analysis with varying distance between cores.}
  \vspace{-10pt}
  \label{fig: thremal_space}
\end{figure}

Although the PE units are placed in a uniform manner, it is observed that the thermal profiling is non-uniform, i.e. the heat flux becomes densely concentrated in certain logic regions and exacerbates the thermal gradients on the logic die.
As shown in Fig. \ref{fig: hotspot_diff} (a), the central area of the logic die typically experiences the highest heat flow due to the highest computation power density, forming the hottest spots of the system.
Such thermal distribution behavior is called \textbf{heat concentration effect}.
We also compare the temperature distribution of logic die in 3D-stacked architecture with 2D architectures under the same power distribution.
As shown in Fig. \ref{fig: hotspot_diff} (b), the 3D architecture exhibits both higher die temperatures and larger temperature differences between hot spots than 2D design.
The heat concentration effect is exacerbated, i.e. larger cross-regional temperature differences, in 3D-stacked architecture due to thermal dissipation constraints.
The maximum temperature gap between hot spots in the 3D architecture is 11.1 $\tccentigrade$, while it is only 2.5 $\tccentigrade$ in 2D architecture.

The exacerbated heat concentration effect of 3D-stacked architectures also leads to more limited performance scalability.
Fig. \ref{fig: t_diff} illustrates the temperature differences between the 3D and 2D architectures across varying cores.
It is observed the 3D-stacked architecture experiences faster temperature rise and wider intra-die temperature disparities with more computation cores.
Specifically, when scaling from 24 to 72 PE cores, the peak temperature of 2D architecture increases by 15.3 $\tccentigrade$, while 3D architecture increases by 23.3 $\tccentigrade$.
As the higher total power intensifies heat accumulation in central regions, the elevated peak temperatures and thermal gradients are more exacerbated across the die, i.e. the 3D architecture shows substantially larger variations (14.8-26.1 $\tccentigrade$).

It is straightforward to improve heat dissipation and lower temperatures by increasing placement distance between cores, while this approach greatly compromises the area utilization. 
Fig. \ref{fig: thremal_space} shows temperature changes with varying X/Y-distance between each core.
When we set the X/Y-distance between each core to half of the original width/height of the core, the peak temperature drops by 10.32 $\tccentigrade$, while the area increases to 2.25$\times$. Hence it is challenging to maintain high performance while preventing overheating within a constrained area.

\begin{figure}[t]
  \centering
  \includegraphics[width=1\linewidth]{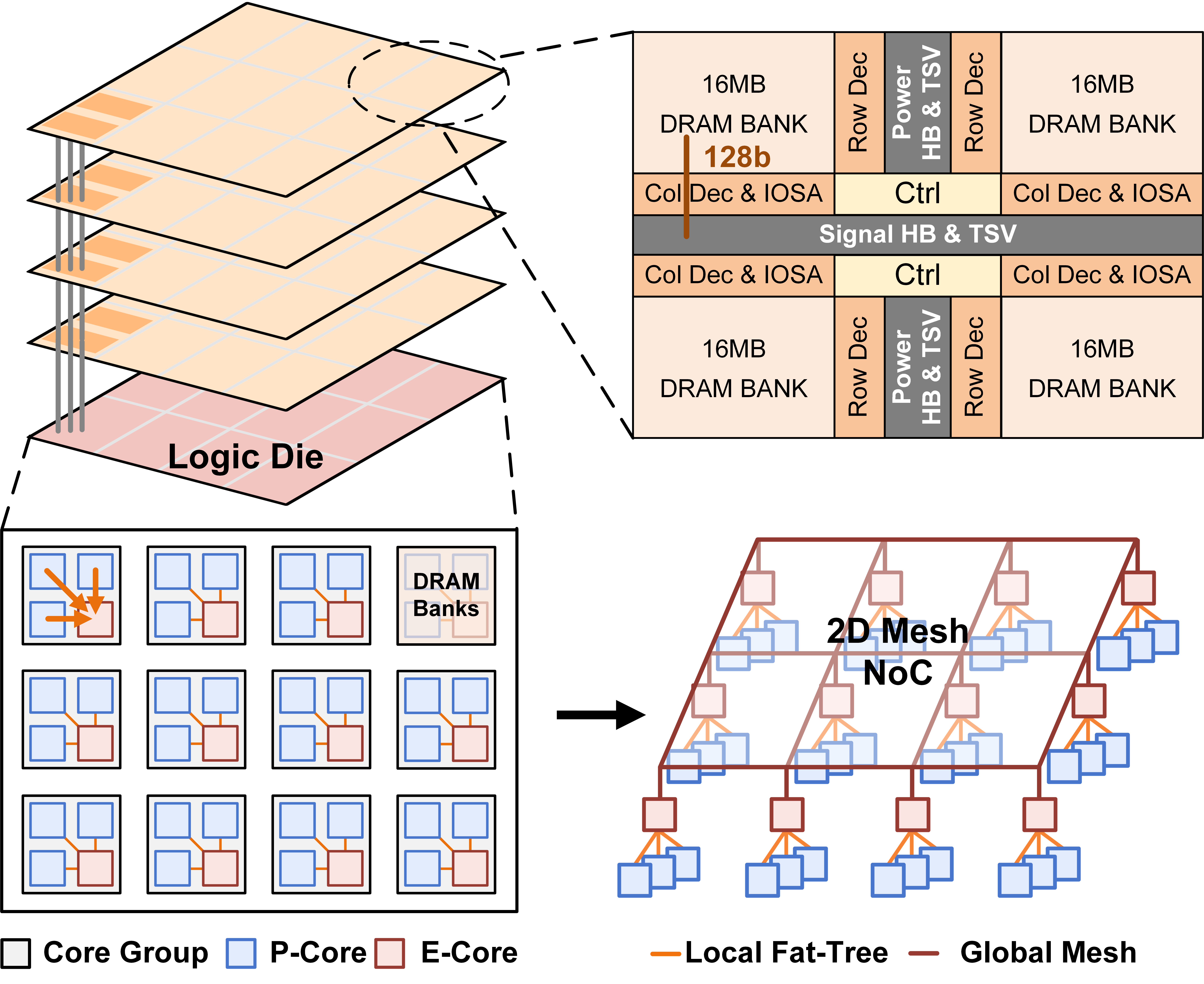}
  \vspace{-15pt}
  \caption{Architecture overview of Tasa, including the stacked DRAM dies and the architecture of logic die. }
  \vspace{-10pt}
  \label{fig: overall_architecture}
\end{figure}

Based on the thermal analysis above, it is evident that 3D-stacked architectures have more challenging thermal issue than conventional 2D design due to the exacerbated heat concentration effect, resulting in \textbf{higher overall temperature}, \textbf{larger temperature gradient}, \textbf{and poorer scalability}.
Therefore, we explore a 3D-stacked engine with a more thermal balanced logic die leveraging heterogeneous cores to ease the heat flow and flatten the temperature distribution.

\section{Tasa Architecture}

\subsection{Overall Architecture}

To address the thermal challenge, we introduce Tasa, a thermal-aware logic die design with two key techniques: thermal-aware heterogeneous architecture and bandwidth sharing scheduling.
Considering the manufacturing capability, we stack four DRAM dies on top of one logic die.
Fig. \ref{fig: overall_architecture} shows the overall architecture of the 48-core Tasa, which can further scale up to 60 and 72 cores with a larger area and more serious thermal issue.
For 48-core Tasa, each stacked DRAM die provides 12 GB memory capacity with 768 DRAM banks, in which each DRAM bank is an independent memory channel with 128 I/Os.
The logic die consists of 6$\times$8 heterogeneous cores, i.e. either P-core or E-core, with each core directly accessing 16 local DRAM bank groups on top of it. By combining thermal awareness with the computational characteristics of LLM operators, we adopt heterogeneous P-cores and E-cores in the logic die for either compute-intensive GEMM operators and memory-intensive GEMV operators.

The logic die adopts a hierarchical NoC to connect heterogeneous cores. The cores are organized into multiple core groups, which contain one E-core and multiple P-cores.
Within a core group, each P-core has a local path connected to the E-core, forming a fat-tree-like NoC, where the E-core acts as the root node and the P-cores act as leaf nodes.
At the global level, all core groups are interconnected through a 2D-mesh NoC, where the E-core acts as the proxy core of its core group to communicate with other core groups.
For LLM inference,  inter-core communication patterns are regular and uniform, dominated by the all-reduce operation.
These characteristics allow the hierarchical NoC to achieve efficient communication with less overhead. 

\subsection{Heterogeneous Core Design}

\begin{figure}[t]
  \centering
  \includegraphics[width=1\linewidth]{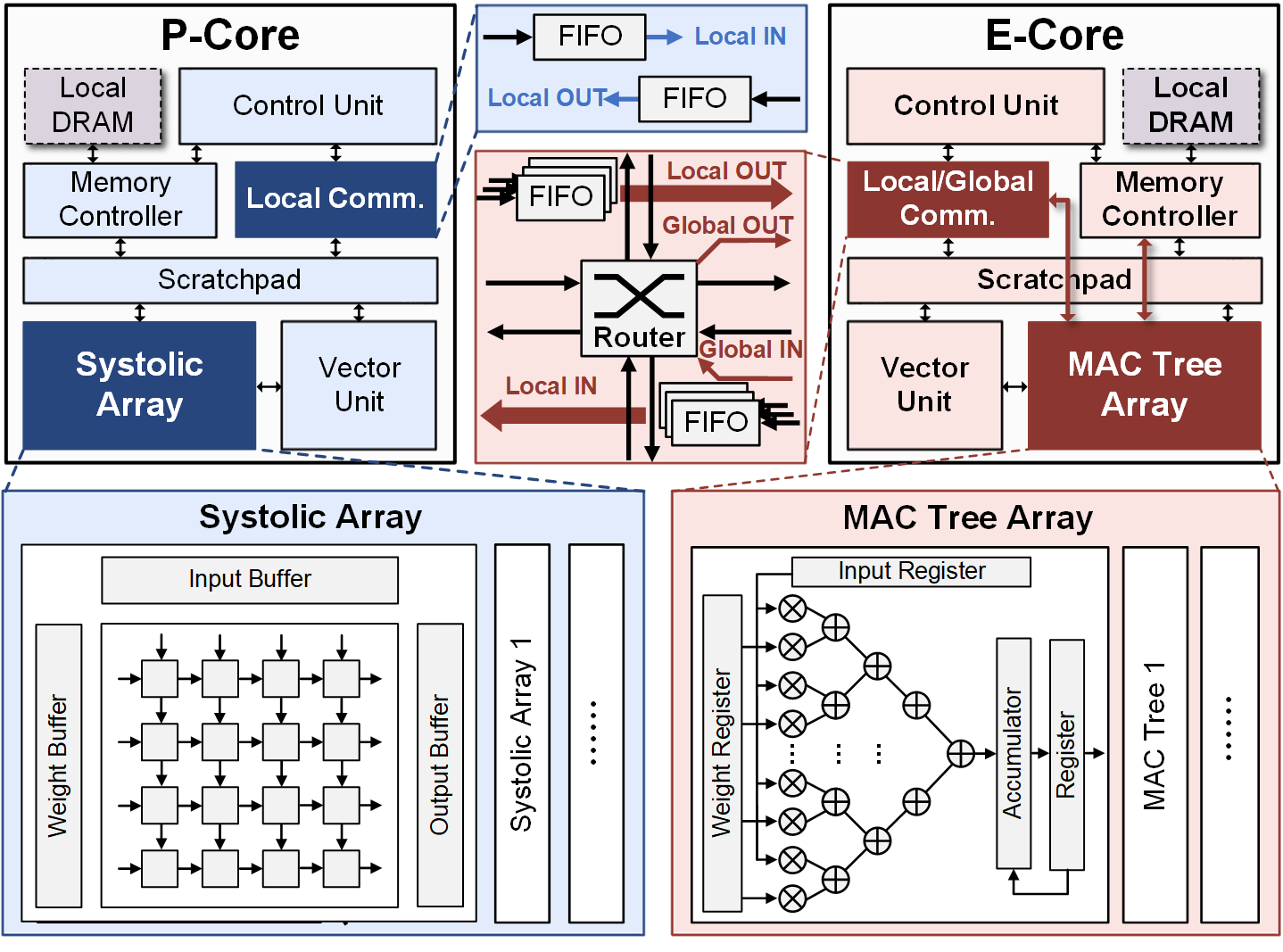}
   \vspace{-20pt}
  \caption{The micro-architecture design of P-core and E-core.}
   \vspace{-15pt}
  \label{fig: core}
\end{figure}

Fig. \ref{fig: core} shows the micro-architecture details of P-core and E-core.
Each core involves a few key components, including control unit, memory controller, scratchpad, communication unit, and an array-based computation engine.
The control unit is primarily responsible for managing computation tasks according to the instructions.
The memory controller and the communication unit manage communication between their local DRAM and other cores, respectively.
The vector units perform non-linear functions and vector operations.
It can independently perform vector operators or act as the post-process unit of the computation engine.
The difference between P-cores and E-cores is the computation array design and the communication unit.

The P-core is designed with the high parallelism systolic array for compute-intensive GEMM operations, e.g. FC layers.
The systolic array reads inputs and weights from the scratchpad, with the result/partial sum written back to the scratchpad or post-processed by the vector unit.
The E-core is optimized for efficiency and utilizes a MAC tree array as a streamlined execution engine for memory-intensive GEMV operations in attention layers.
Each MAC tree receives different weights, while the input is broadcast to all the MAC trees.
The MAC tree array supports flexible data access through multiple pathways, which can not only read data from the local buffer but also bypass local buffers to read data directly from the local DRAM and the communication unit.
This architecture optimizes attention operations by eliminating redundant data across the memory hierarchy.

Each P-core is equipped with a router-less NoC communication unit, providing a direct and dedicated connection to its root E-core.
In contrast, each E-core incorporates an additional router-based NoC communication unit to manage inter-group global communication.
The E-core can simultaneously communicate with all its leaf P-cores through the dedicated intra-group path for efficient bandwidth sharing, with NoC bandwidth matched to DRAM bandwidth in each core.

\subsection{Heterogeneous Architecture Exploration }

The heterogeneous designs with P-cores and E-cores introduce a design exploration requirement of the core ratio and floorplan topologies for performance and temperature distribution optimizations.
With increasing the number of E-cores, the heat dissipation is improved while the peak performance is reduced.
More P-cores will enhance the performance, but leading to higher temperature and imposing more frequency throttling to degrade the performance gain.
Therefore, it is crucial to select an optimal ratio between P-cores and E-cores for balancing performance and thermal limits. Meanwhile, the strategic floorplanning of these heterogeneous cores is also vital to mitigate hot spots since leveraging architectural heterogeneity can distribute the heat flow more evenly across the die and maximize effective performance.

It is worth to mention the design space for the core ratio is also constrained by the core organization and the design space of floorplan topology is extremely large.
Different core ratios result in distinct design spaces of floorplan topologies, each affecting heat distribution on the logic die in unique ways.
The distinction between various floorplans lies in the spatial clustering density of E-cores. 
To optimize the design space search, we heuristically place a subset of E-cores in the central region, followed by the thermal analysis on different floorplans with varying spatial clustering density of the remaining E-core.
In this work, we particularly examine ratios of 3:1 and 5:1 due to the constraints of the core organization.

\section{Bandwidth-Sharing for LLM Inference}

\subsection{Dataflow and Mapping}

\begin{figure}[t]
  \centering
  \includegraphics[width=0.95\linewidth]{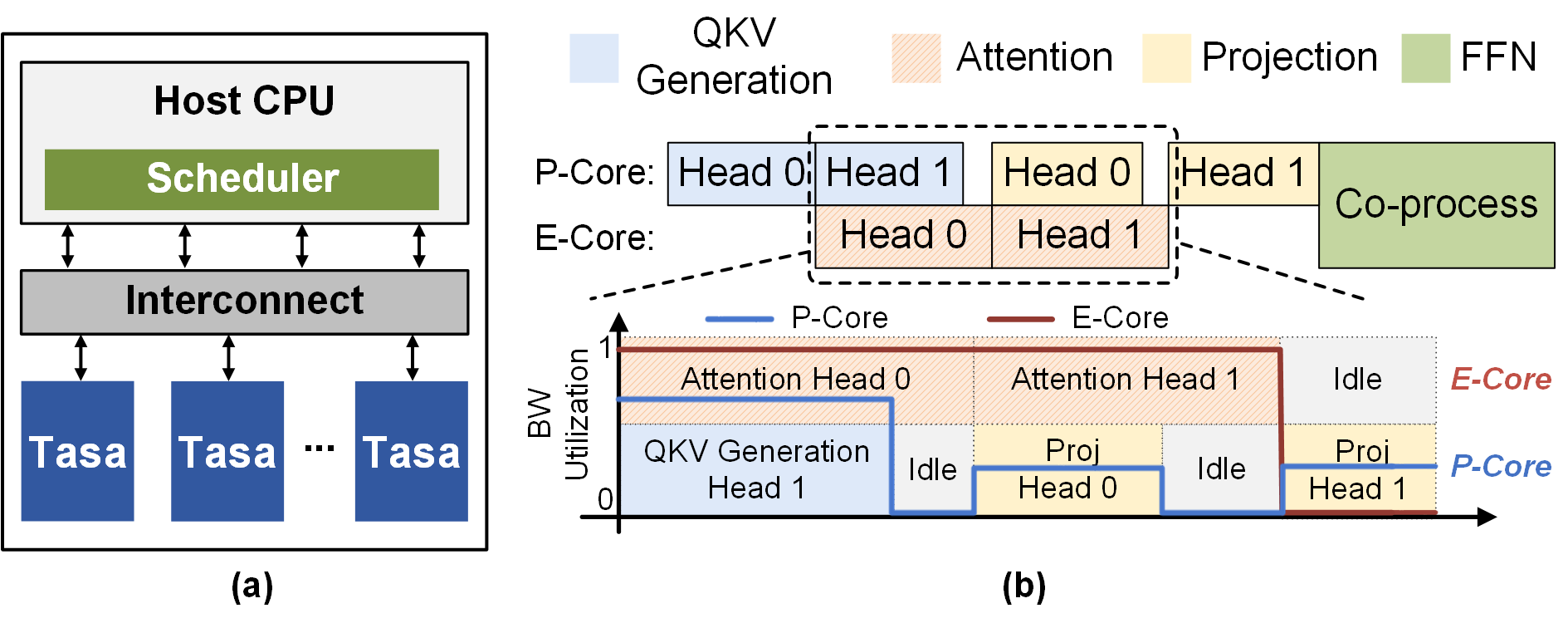}
  \vspace{-10pt}
  \caption{(a) Inference platform with Tasa. (b) dataflow for Tasa with low bandwidth utilization.}
  \vspace{-15pt}
  \label{fig: platform}
\end{figure}

Fig. \ref{fig: platform} (a) shows the platform architecture using Tasa for LLM  inference. To accommodate to the LLM model size, multiple 3D-stacked Tasa architecture can be integrated as a inference platform. The interconnect between Tasa devices and host CPU adopts standard interconnect such as PCIe.
Fig. \ref{fig: platform} (b) top shows our dataflow across Tasa. Simiar as the dataflow in heterogeneous systems \cite{attacc_2024}, in our case, P-cores process computation-intensive FC layers in QKV generation and projection while the memory-intensive attention layers are offloaded to E-cores.
It adopts the head-level pipeline, in which the FC layers for QKV generation and projection are pipelined with attention layers to reduce overall latency. 
To ensure high parallelism, tensor parallelism is implemented across the devices to enhance the performance.
Within each device, the model parameters are evenly partitioned across cores at the tensor level to avoid remote memory access.
The weight for QKV generation and projection is distributed within P-cores, while the weight for FFN is distributed in both P-cores and E-cores.
The KV cache is all stored in E-cores.

Tasa's architecture offers improved thermal characteristics, while its dataflow and mapping need carefully optimizations for well hardware utilization.
As illustrated Fig. \ref{fig: platform} (b) bottom, P-cores may show underutilization during head-level pipeline execution since the bandwidth of 3D-stacked DRAM is distributed across banks and P-cores contain the most bandwidth.
Consequently, heterogeneous architectures require optimized dataflow scheduling to maximize hardware utilization.

\subsection{Bandwidth Sharing for Tasa}

To fully exploit the potential of Tasa, we also introduce a bandwidth sharing technique to improve the system bandwidth utilization.
During the head-level pipeline, the P-cores are either executing FC layers or remain idle for the attention layer results.
When executing FC layers, the P-cores may not fully utilize the bandwidth due to the data reuse in batching.
This provides an opportunity to share the unused bandwidth with the E-cores through the local fat-tree NoC.

To leverage the shared bandwidth, KV cache should be appropriately allocated across both P-cores and E-cores, as the latency of one head in the attention layer for one request is $max(KV_{P}/BW_{shared}, KV_{E}/BW_{core})$, in which $KV_{P}$ and $KV_{E}$ refer to KV cache in P-cores and E-cores, respectively.
Thus, KV cache is required to be allocated between them according to the ratio of $BW_{shared}$ and $BW_{core}$ to minimize the latency, in which $BW_{shared}$ is determined by the state of P-cores and batch size.
$BW_{shared}$ can be allocated as the total bandwidth of P-cores when they are idle or depends on the bandwidth utilization in P-cores when executing FC layers with varying batch size.
However, the dynamic variations in KV cache and batch size make it challenging to maintain optimal allocation ratios during inference.

\begin{figure}[t]
  \centering
  \includegraphics[width=0.95\linewidth]{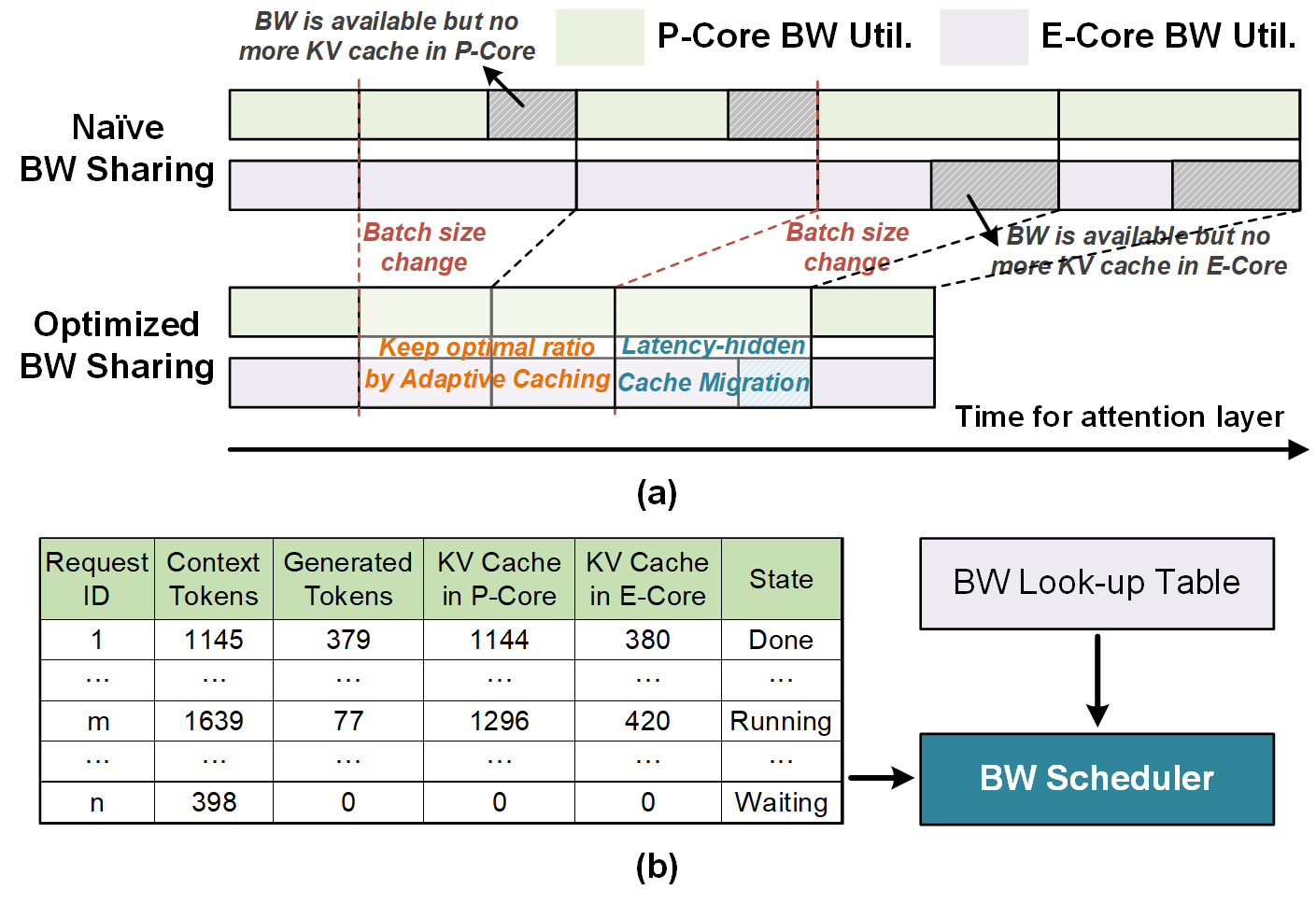}
  \vspace{-10pt}
  \caption{(a) Comparison between naive bandwidth sharing and our optimized bandwidth sharing (b) The flow of bandwidth sharing scheduling.}
  \vspace{-10pt}
  \label{fig: scheduling}
\end{figure}

To effectively leverage bandwidth sharing, we implement adaptive caching to efficiently execute the attention layer in the current iteration, along with hiding the latency of reallocation to maintain an optimal allocation ratio for subsequent iterations, as shown in Fig. \ref{fig: scheduling} (a) .
These are built on two observations.
One is that the batch size changes only slightly with each iteration.
The other is that the shared bandwidth is typically constant for a given batch size.
Fig. \ref{fig: scheduling} (b) shows the overall scheduling flow. 
During the inference, we record the KV cache allocation for each request in the host CPU. A batch-size-indexed lookup table (LUT) is maintained for the runtime bandwidth utilization. 
By monitoring the change of batch size, we get the optimal allocation ratio based on the bandwidth utilization LUT and schedule KV cache migration at each iteration.

\begin{figure}[t]
  \centering
  \includegraphics[width=0.95\linewidth]{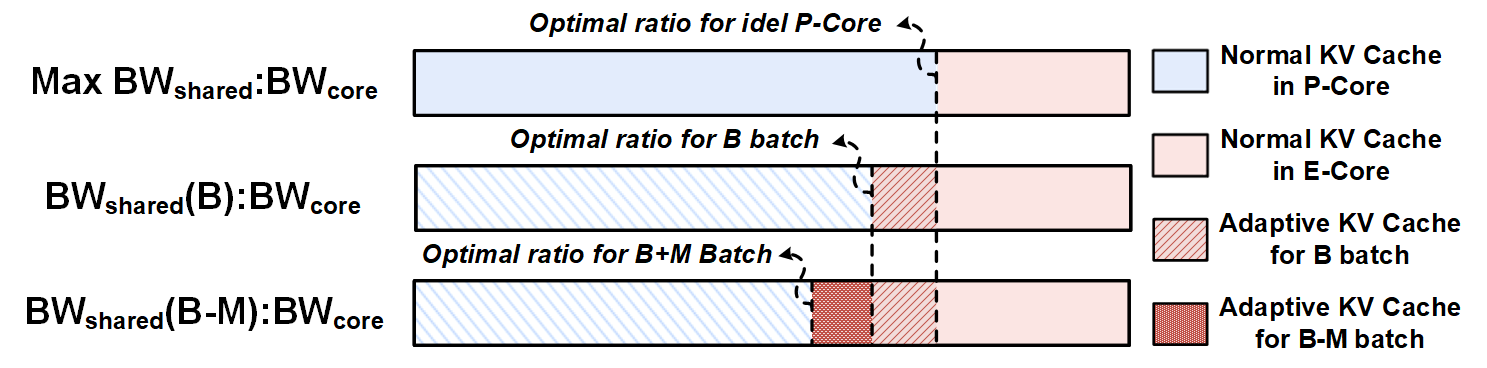}
  \vspace{-10pt}
  \caption{KV Cache partition under our bandwidth sharing technique.}
  \vspace{-15pt}
  \label{fig: dataflow}
\end{figure}

Fig. \ref{fig: dataflow} illustrates how the KV cache is adaptive partitioned with bandwidth sharing.
We first partition KV cache based on the total bandwidth of P-cores, which ensures that for the bandwidth of idle P-core can be fully utilized.
Then we allocate the remaining KV cache to E-cores. 
Guided by the bandwidth utilization LUT, extra KV cache is allocated to E-cores, targeting $BW_{share}(B)$ to ensure optimal performance when P-cores are executing FC layers.
Finally, we place additional KV cache in E-cores based on $BW_{share}(B-M)$ to prevent performance loss due to the batch size change. $M$ is a configurable parameter tuned for each LLM workload.
This bandwidth sharing has sufficient flexibility for optimal KV cache allocation under various P-core states and batch sizes.

Once the batch size changes, KV cache reallocation will be invoked to restore the optimal allocation ratio for the following iteration.
When the batch size increases, the current adaptive KV cache remains sufficient to accommodate the change, allowing the optimal allocation ratio to be preserved. If the batch size decreases, E-cores will gather additional KV cache to meet the adaptive caching requirements. Since E-cores inherently receive KV cache from P-cores to execute attention layers, this extra KV cache can be migrated to them at no additional overhead, replenishing their adaptive cache to handle batch size fluctuations. After migration, the newly generated K and V are assigned in a way that maintains the existing optimal KV cache allocation ratio.

\begin{table}[]
\vspace{-10pt}
\centering
\caption{\textbf{Hardware Specification}}
\label{tab: hardware}
\renewcommand{\arraystretch}{1.1}
\begin{tabular}{cccc}
\hline
\multicolumn{4}{c}{\textbf{Logic Die}} \\ \hline
\multicolumn{1}{c|}{\# Core} & \multicolumn{1}{c|}{48/60/72} & \multicolumn{1}{c|}{SRAM/Core} & 128 KB \\
\multicolumn{1}{c|}{NoC Flit Width} & \multicolumn{1}{c|}{128 B} & \multicolumn{1}{c|}{Base Frequency} & 1 GHz \\ \hline
\multicolumn{2}{c|}{FC Core} & \multicolumn{2}{c}{Attention Core} \\ \hline
\multicolumn{1}{c|}{Systolic Array/Core} & \multicolumn{1}{c|}{2} & \multicolumn{1}{c|}{MAC Tree Array/Core} & 12 \\
\multicolumn{1}{c|}{Vector Unit/Core} & \multicolumn{1}{c|}{2} & \multicolumn{1}{c|}{Vector Unit/Core} & 2 \\
\multicolumn{1}{c|}{Systolic Array Size} & \multicolumn{1}{c|}{32 $\times$ 32} & \multicolumn{1}{c|}{MAC Tree Size} & 32 $\times$ 1 \\
\multicolumn{1}{c|}{Vector Unit Size} & \multicolumn{1}{c|}{32 $\times$ 1} & \multicolumn{1}{c|}{Vector Unit Size} & 32 $\times$ 1 \\ \hline
\multicolumn{4}{c}{\textbf{DRAM Die}} \\ \hline
\multicolumn{1}{c|}{Die/Stack} & \multicolumn{1}{c|}{4} & \multicolumn{1}{c|}{Capacity/Stack} & 48 GB \\
\multicolumn{1}{c|}{Banks/Core} & \multicolumn{1}{c|}{16} & \multicolumn{1}{c|}{Capacity/Die} & 12 GB \\
\multicolumn{1}{c|}{Banks/Die} & \multicolumn{1}{c|}{768} & \multicolumn{1}{c|}{Capacity/Bank} & 16 MB \\ \hline
\multicolumn{1}{c|}{Area} & \multicolumn{3}{c}{20 mm $\times$ 24 mm} \\  \hline
\multicolumn{1}{c|}{Data rate} & \multicolumn{3}{c}{128 IO/Bank, 500 MHz,  6 TB/Stack} \\  \hline
\multicolumn{1}{c|}{Timing Parameter} & \multicolumn{3}{c}{\begin{tabular}[c]{@{}c@{}}tRC = 45.3 ns, tRCD = 16.4 ns, tCL = 1.7 ns,\\ tRAS = 33.6 ns, tRP = 11.7 ns\end{tabular}} \\  \hline
\multicolumn{1}{c|}{Refresh Period} & \multicolumn{3}{c}{\begin{tabular}[c]{@{}c@{}}$\leq$ 85 $\tccentigrade$ 32 ms, $\leq$ 95 $\tccentigrade$ 16 ms\\ $\leq$ 105 $\tccentigrade$ 8 ms, $\leq$ 115 $\tccentigrade$ 4 ms\end{tabular}} \\  \hline
\end{tabular}
\end{table}

\begin{table}[]
\vspace{-10pt}
\caption{\textbf{Thermal Simluation Parameters}}
\label{tab: thermal}
\renewcommand{\arraystretch}{1.1}
\centering
\scalebox{1.1}{
\begin{tabular}{ccc}
\hline
\textbf{}       & \textbf{TIM}             & \textbf{TSV}              \\ \hline
Heat Capacity   & $4 \times 10^6 J/m^3K$   & $4 \times 10^6 J/m^3K$ \\
Resistivity     & $0.25 mK/W$              & $0.0058 mK/W$               \\
Thickness       & $0.02 mm$                & $0.02 mm$                 \\ \hline
\textbf{}       & \textbf{Logic Silicon}   & \textbf{DRAM Silicon}     \\ \hline
Heat Capacity   & $1.75 \times 10^6 J/m^3K$   & $1.75 \times 10^6 J/m^3K$ \\
Resistivity     & $0.01 mK/W$              & $0.01 mK/W$               \\
Thickness       & $0.015 mm$               & $0.15 mm$                 \\ \hline
\multicolumn{3}{c}{\textbf{Heat Sink}}                                 \\ \hline
\multicolumn{2}{c}{Convection Capacitance} & $140.4 J/K$               \\
\multicolumn{2}{c}{Convection Resistance}  & $0.1 K/W$                 \\
\multicolumn{2}{c}{Thermal Conductivity}   & $400 W/mK$                \\ \hline
\end{tabular}
}
\vspace{-15pt}
\end{table}

\section{Evaluation}
\subsection{Experimental Setup}
\noindent \textbf{Workload:} We set LLaMA-65B \cite{llama-65B}, GPT-3 66B \cite{gpt} as the target model with FP16/BF16 data type.
We conduct our evaluation using real-world conversational requests sampled from ShareGPT, which is a set of conversations taken from the real-world user log of ChatGPT.
We evaluate the performance of stable batch sizes by dynamically adding the next request whenever an inference completes.
For varying queries per second (QPS) situations, we generate request arrivals through a Poisson process to simulate realistic traffic patterns.

\noindent \textbf{Hardware specifications:}
The hardware configuration of the Tasa device is listed in Table \ref{tab: hardware}.
The threshold temperature constraint is set to 85 $\tccentigrade$.
Once the maximum temperature exceeds the threshold temperature, DVFS is triggered to prevent thermal runaway.
We evaluate the performance for different target models with the core ratio of 3:1 and 5:1.

\noindent \textbf{Baseline:}
For evaluation, we compare our Tasa inference platform (Fig.~\ref{fig: platform}) with three baselines: (a) \textit{A100-only:} a computing platform with only 8 NVIDIA A100 40GB GPUs;
(b) \textit{A100+AttAcc:} a integrated platform with 8 NVIDIA A100 GPUs and AttAcc PIM, which is the state-of-the-art LLM accelerator;
(c) \textit{Homogeneous 3D-stacked accelerator (Homo-3D):} a computing platform with 8 homogeneous 3D-stacked accelerators with identical P-cores, which has the same configuration of Tasa.
Homo-3D adopts the same threshold temperature and DVFS as our heterogeneous system.

\noindent \textbf{Simulation:}
We developed a simulator framework by a combination of AttAcc \cite{attacc_2024}, Timeloop \cite{timeloop}, Accelergy \cite{accelergy}, Ramulator 2.0 \cite{ramulator2_2023}, and HotSpot 6.0 \cite{HotSpot6_2015} for evaluations.
To evaluate the performance, we assess the performance of each operator using Timeloop, with DRAM accesses offloaded to Ramulator.
To evaluate the area and energy, we extend Accelergy for 7nm process (ASAP7) \cite{asap7_2016} and the 3D-stacked DRAM evaluation.
we modified FinCACTI with the published information of SRAM to model 7nm SRAM.
The energy consumption of stacked DRAM refers to the latest data from \cite{sedram2}, while the area of the memory controller is estimated from \cite{alibaba_3d}.
Finally, we augment the AttAcc framework with the above results for end-to-end inference evaluation and power trace generation.
For thermal evaluation, we utilize HotSpot 6.0 to analyze the temperature with the floorplan and power traces, with detailed configuration in Table. \ref{tab: thermal}.

\begin{figure}[t]
  \centering
  \includegraphics[width=1\linewidth]{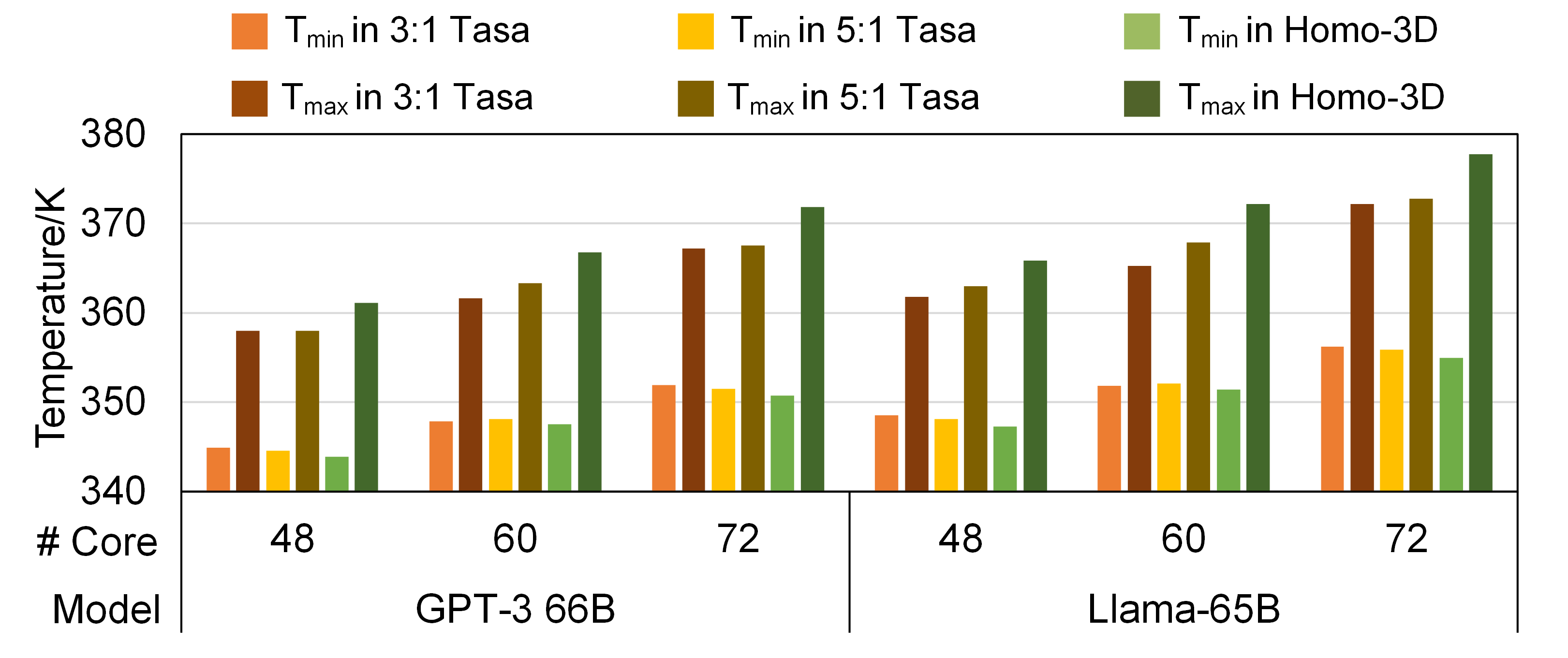}
  \vspace{-20pt}
  \caption{Thermal improvement of Tasa for LLM inferences.}
  \vspace{-15pt}
  \label{fig: e2e_t}
\end{figure}

\subsection{Thermal Improvement of Tasa}
Fig. \ref{fig: e2e_t} shows the steady temperature of various core counts and LLM models for a batch size of 32 and ($L_{in}, L_{out}$) of (512, 512).
Tasa achieves average peak temperature reductions of 3.96 $\tccentigrade$, 6.10 $\tccentigrade$, and 5.99 $\tccentigrade$ for 48, 60, and 72 cores compared to Homo-3D, with minimal impact on the lowest temperature.
Since our the heterogeneous architecture rebalance power distribution and the heat flow, Tasa can effectively suppress server hot spot formation to reduce the peak temperature.
The thermal benefits of Tasa become more pronounced for larger core counts, as the overall power increased with core counts, which leads to higher temperature and more unbalanced heat distribution.

\begin{figure}[t]
  \centering
  \includegraphics[width=1\linewidth]{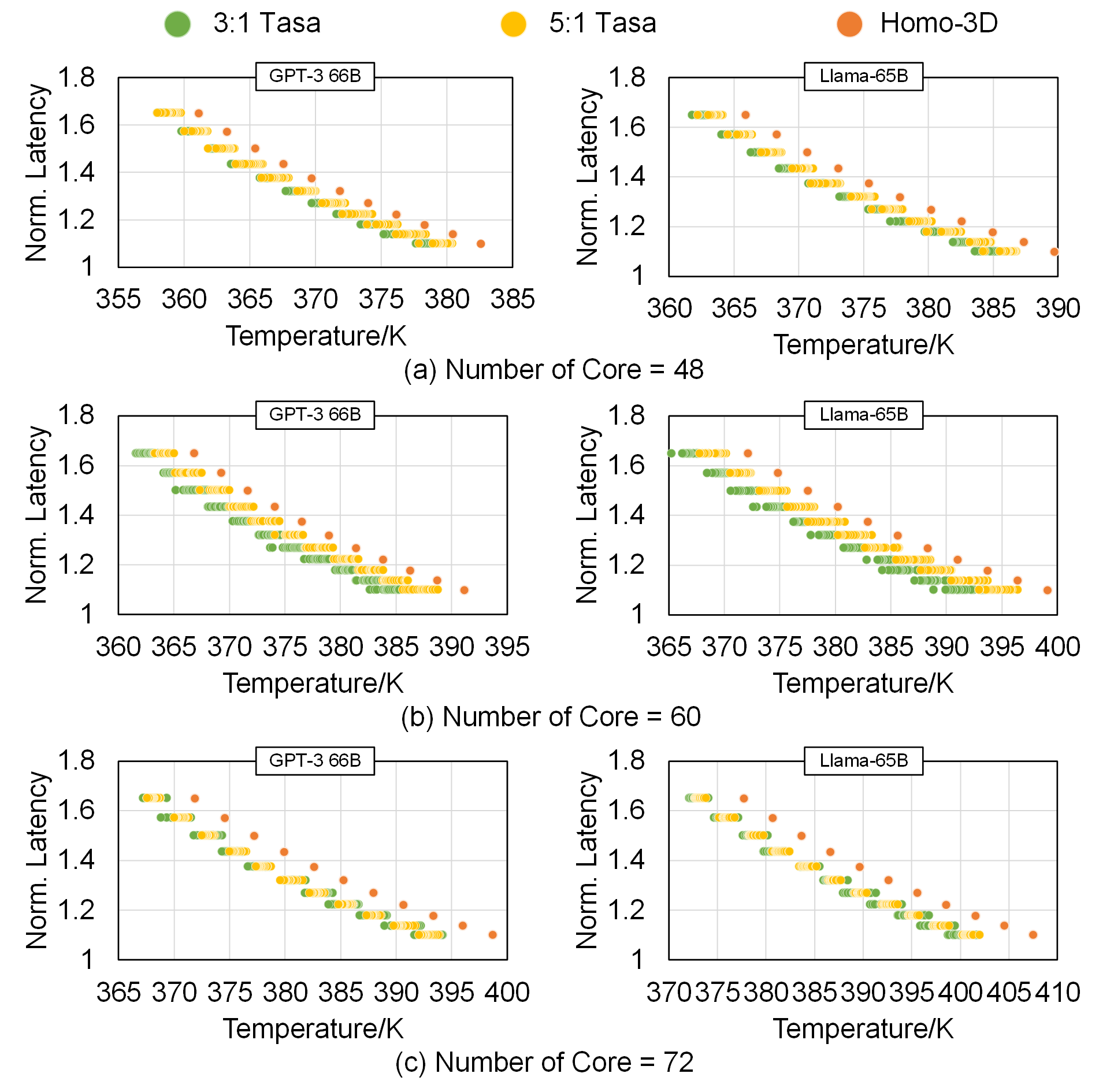}
  \vspace{-20pt}
  \caption{Heterogeneous layout design space visualization.}
  \vspace{-15pt}
  \label{fig: dse}
\end{figure}

Fig. \ref{fig: dse} visualizes the design space of the heterogeneous floorplan under different core counts for the same workload configuration of Fig. \ref{fig: e2e_t}.
We vary the base frequency of Homo-3D to analyze its thermal characteristics, while applying frequency boost to Tasa for performance-equivalent comparison.
Tasa achieves average peak temperature reduction of 
3.63-5.55 $\tccentigrade$, 5.73-9.37 $\tccentigrade$, 5.12-7.91 
 $\tccentigrade$ for 48, 60, and 72 cores. 
The results clearly show that Tasa extends the latency-temperature Pareto frontier, enabling a lower temperature for a given performance, with greater thermal benefits as both core counts and frequency increase.
Overall, Tasa presents better thermal performance and scalability than Homo-3D under all conditions, in which the core ratio of 3:1 outperforms the core ratio of 5:1, as it has more E-cores, providing more thermal optimization opportunities.

\begin{figure*}[t]
  \centering
  \includegraphics[width=1\linewidth]{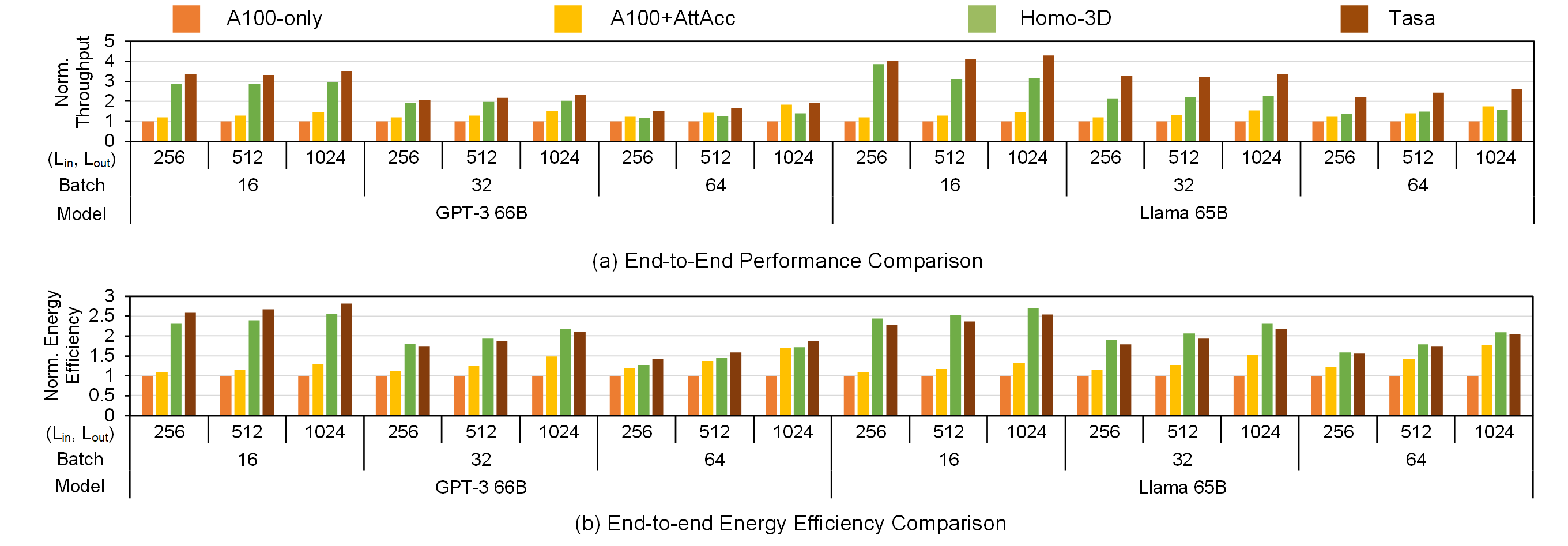}
  \vspace{-20pt}
  \caption{The normalized end-to-end throughput and energy efficiency comparison for LLM inferences.}
  \vspace{-15pt}
  \label{fig: e2e}
\end{figure*}

\begin{figure}[t]
  \centering
  \includegraphics[width=1\linewidth]{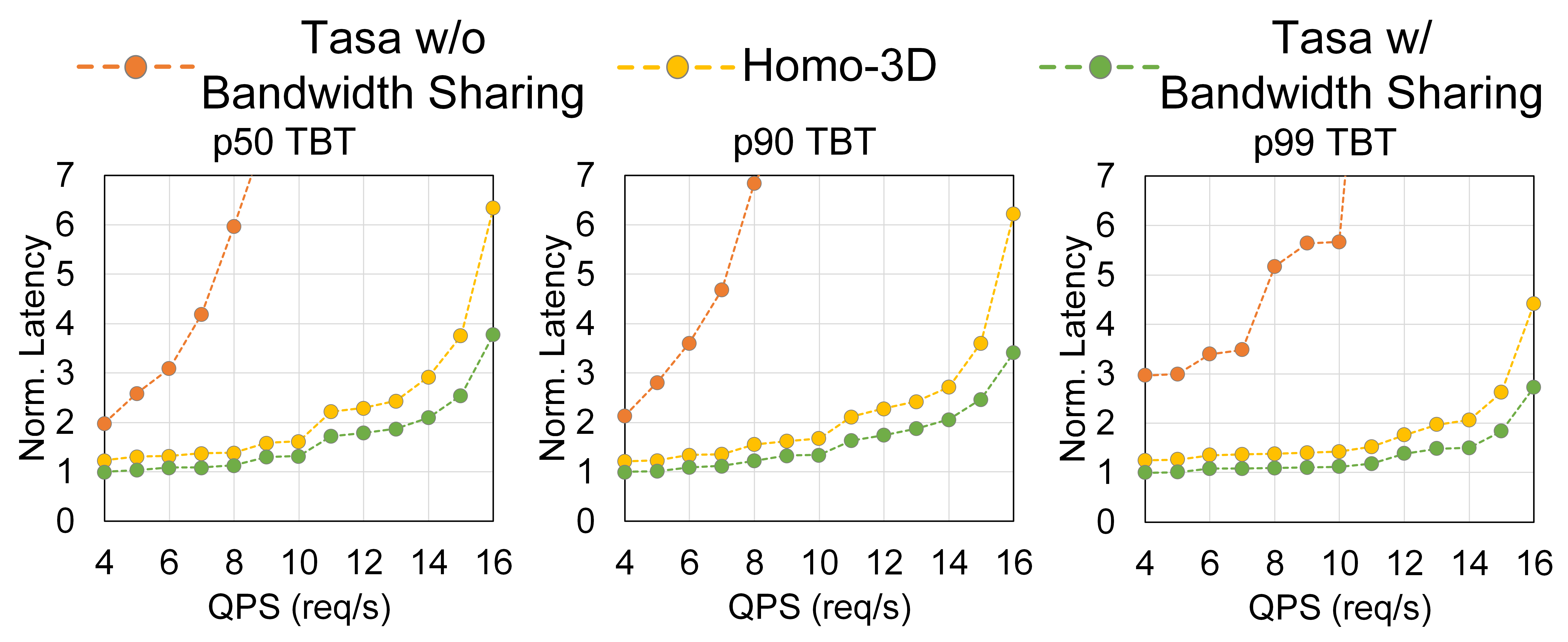}
  \vspace{-20pt}
  \caption{The normalized throughput and energy.}
  \vspace{-15pt}
  \label{fig: qps}
\end{figure}

\subsection{Performance Improvement of Tasa}
We evaluate the performance of Tasa using a 48-core configuration, compared with the three baselines for various batch sizes and ($L_{in}, L_{out}$) configurations on two models.
Fig. \ref{fig: e2e}(a) shows the comparison of normalized throughput.
Tasa demonstrates significant performance advantages across three baselines.
Compared to the A100-only baseline, Tasa delivers a 2.85$\times$ average speedup by exploiting its 3D-stacked architecture's superior memory bandwidth, while the A100 remains constrained by off-chip bandwidth limitations.
Against A100+AttAcc, Tasa achieves a 2.12$\times$ performance gain.
Although A100+AttAcc benefits from PIM's high internal bandwidth for attention layers, its FC layer performance still suffers from off-chip bandwidth constraints.
In contrast, Tasa enables consistent performance improvements for both FC and attention layers, with particularly significant gains (up to 3.35×) at smaller batch sizes.
Furthermore, Tasa outperforms Homo-3D by 1.33$\times$ on average.
Tasa converts thermal headroom into boosted frequency while leveraging bandwidth sharing, enhancing overall system performance.
Notably, the speedup gains grow with larger batch sizes.
As the computational intensity of FC layers scales with batch size, the increased power density exacerbates hot spot formation.
As Tasa exhibits enhanced hot spot suppression capabilities precisely under high power density scenarios, more temperature reduction is obtained compared to low-power scenarios.

Fig. \ref{fig: e2e}(b) shows the comparison of normalized energy efficiency.
Tasa attains average energy efficiency improvement of 2.07 $\times$ and 1.61 $\times$ over A100-only and A100+AttAcc, respectively, primarily due to the inherently lower memory access energy of 3D-stacked technology.
The improvement is particularly significant at smaller batch sizes, where model weight transfers dominate power consumption.
Additionally, Tasa achieves energy efficiency comparable to Homo-3D.
This is primarily because the E-cores enhance the energy efficiency of attention layers, while the bandwidth sharing scheduling introduces negligible NoC power consumption.

Fig. \ref{fig: qps} shows the latency of the Homo-3D and Tasa running GPT-3 66B under varying QPS.
Without bandwidth sharing, Tasa presents larger latency than Homo-3D, which increases rapidly with QPS.
Adopting bandwidth sharing, Tasa can fully utilize the bandwidth and improve the overall bandwidth utilization, always outperforming Homo-3D.
As QPS increases, the system is typically running with a higher batch size, where the latency of the attention layers becomes more dominant, and P-cores present lower bandwidth utilization.
Thus, there is more ample room for bandwidth sharing scheduling to reduce the latency at high QPS.
Overall, the bandwidth sharing scheduling improves the hardware utilization of the heterogeneous architecture, further enhancing the performance of Tasa.

\section{Discussions}

Prior research has widely studied thermal optimization for 3D-stacked architecture, such as DTM and thermal-aware floorplanning.
Traditional DVFS lowers the frequency of all cores to reduce temperature, but this approach decreases overall performance more than it needs to, as some cores are far from reaching their thermal limits.
However, adopting finer-grained DVFS introduces significant overhead and synchronization challenges.
Another common approach is workload migration, which migrates the workload from hot cores to cool cores.
Due to the distributed weights across DRAM banks, the weights in the hot cores will be heavily accessed by other cores, resulting in NoC congestion and workload unbalance.

Compared with common DTMs, such as DVFS and workload migration, Tasa alleviate thermal issue at architecture level for LLM inference, which can effectively reduce the temperature and converter the thermal benefits to performance gains.
In contrast, DTMs can only trade performance loss for the temperature reduction. 
For example, DVFS lowers the operation frequency to reduce temperature while workload migration shut down the hot cores and migrates the workload to cool cores.

Furthermore, other optimization techniques such as thermal-aware floorplanning are orthogonal to our approach.
For example, prior research has investigated floorplanning method for many-core 3D systems \cite{floorplan1} to refine both horizontal and vertical layouts, which can also be applied to augment Tasa.
The fundamental contributions of Tasa lie in introducing heterogeneous architecture optimized for LLM inference and demonstrating its superiority over homogeneous designs in both thermal and inference metrics through comprehensive evaluations, which enable further research on floorplanning optimization for heterogeneous 3D-stacked architectures.

\section{Conclusion}
In this work, we propose Tasa, a thermal-aware heterogeneous architecture with bandwidth sharing for LLM inference.
We offer a comprehensive thermal analysis of 3D-stacked architecture for LLM inference and revealing the intensified heat concentration effect is the primary contributor to thermal challenges.
We design a heterogeneous 3D-stacked architecture with performance cores and efficiency cores to balance the heat distribution and reduce the temperature.
Dataflow optimization with bandwidth sharing is adopted to improve the hardware utilization.
Extensive evaluations demonstrate that Tasa effectively mitigates thermal issues, achieving 
speedups of 2.85×, 2.21×, and 1.33× over GPU, PIM-based accelerator, and the homogeneous 3D-stacked baselines.

\section*{Acknowledgement}
This work was supported in part by NSFC No. 92464202.


\bibliographystyle{ieeetr}
\bibliography{refs}

\end{document}